%% file: main.tex
\documentclass[twocolumn,showpacs,aps,prd,superscriptaddress]{revtex4-1}
\usepackage[english]{babel}
\usepackage[utf8]{inputenc}

\usepackage[]{lineno}
\input{preamble}
\pdfoutput=1

\begin{document}
\title{Study of the $h_c$($1^1P_1$) meson via $\psi(2S) \rightarrow \pi^0 h_c$ decays at BESIII}

%-----------------------------
\author{\small \center M.~Ablikim$^{1}$, M.~N.~Achasov$^{10,b}$, P.~Adlarson$^{68}$, S. ~Ahmed$^{14}$, M.~Albrecht$^{4}$, R.~Aliberti$^{28}$, A.~Amoroso$^{67A,67C}$, M.~R.~An$^{32}$, Q.~An$^{64,50}$, X.~H.~Bai$^{58}$
% ...
\\
\vspace{0.2cm}
(BESIII Collaboration)\\
\vspace{0.2cm} {\it
$^{1}$ Institute of High Energy Physics, Beijing 100049, People's Republic of China\\
% ...
$^{4}$ Bochum Ruhr-University, D-44780 Bochum, Germany\\
% ...
$^{10}$ G.I. Budker Institute of Nuclear Physics SB RAS (BINP), Novosibirsk 630090, Russia\\
% ...
$^{14}$ Helmholtz Institute Mainz, Staudinger Weg 18, D-55099 Mainz, Germany\\
% ...
$^{28}$ Johannes Gutenberg University of Mainz, Johann-Joachim-Becher-Weg 45, D-55099 Mainz, Germany\\
% ...
$^{32}$ Liaoning Normal University, Dalian 116029, People's Republic of China\\
% ...
$^{50}$ State Key Laboratory of Particle Detection and Electronics, Beijing 100049, Hefei 230026, People's Republic of China\\
% ...
$^{58}$ University of Jinan, Jinan 250022, People's Republic of China\\
% ...
$^{64}$ University of Science and Technology of China, Hefei 230026, People's Republic of China\\
% ...
$^{67}$ University of Turin and INFN, (A)University of Turin, I-10125, Turin, Italy; (B)University of Eastern Piedmont, I-15121, Alessandria, Italy; (C)INFN, I-10125, Turin, Italy\\
$^{68}$ Uppsala University, Box 516, SE-75120 Uppsala, Sweden\\
% ...
$^{b}$ Also at the Novosibirsk State University, Novosibirsk, 630090, Russia\\
% ...
}}

\date{July 29, 2022} 

\begin{abstract}
Using 448 million $\psi(2S)$ events, the spin-singlet $P$-wave charmonium state $h_c$(1$^1P_1$) is studied via the $\psi(2S) \rightarrow \pi^0 h_c$ decay followed by the $h_c \to \gamma \eta_c$ transition. The branching fractions are measured to be $\mathcal{B}_{\rm Inc}(\psi(2S) \rightarrow \pi^0h_c) \times \mathcal{B}_{\rm Tag}(h_c \rightarrow \gamma \eta_c) = (4.22 ^{+0.27}_{-0.26} \pm 0.19) \times  10^{-4}$ , $\mathcal{B}_{\rm Inc}(\psi(2S) \rightarrow \pi^0h_c) = (7.32 \pm 0.34 \pm 0.41) \times 10^{-4}$, and $\mathcal{B}_{\rm Tag}(h_c \rightarrow \gamma \eta_c) = (57.66 ^{+3.62}_{-3.50} \pm 0.58)\%$, where the uncertainties are statistical and systematic, respectively. The $h_c$(1$^1P_1$) mass and width are determined to be $M = (3525.32 \pm 0.06 \pm 0.15)~\rm{MeV}/c^{\rm{2}}$ and ${\mathit \Gamma} = (0.78 ^{+0.27}_{-0.24} \pm 0.12)~\rm{MeV}$. Using the center of gravity mass of the three $\chi_{cJ}$(1$^3 P_J$) mesons ($M(\rm c.o.g.)$), the $1P$ hyperfine mass splitting is estimated to be $\Delta_{\rm hyp} =  M(h_c) - M(\rm c.o.g.) = (0.03 \pm 0.06 \pm 0.15)  ~ \rm{MeV}/c^{\rm{2}}$, which is consistent with the expectation that the $1P$ hyperfine splitting is zero at the lowest-order. \\

\end{abstract}

\keywords{BESIII, charmonium spectroscopy, QCD}

\maketitle

%\linenumbers
\input{sections/section01.tex}
\input{sections/section02.tex}
\input{sections/section03.tex}

\input{sections/section04.tex}
\input{sections/section05.tex}
\input{sections/acknowledgements.tex}

\nocite{*}
\bibliographystyle{apsrev4-1} 
\bibliography{mybib}

\end{document}

%% file: preamble.tex
\usepackage{graphicx}
\usepackage{dcolumn}
\usepackage{amsmath}
\usepackage{epsfig}
\usepackage{subfloat}
\usepackage{colordvi}
\usepackage{hhline}
\usepackage{multirow}
\usepackage{subfigure}
\usepackage{rotating}
\usepackage{slashbox}
\usepackage{hyperref}
\usepackage{xcolor}
\usepackage{lineno}
\usepackage{url}

\usepackage{mathtools}
\usepackage{physics}
\usepackage{xcolor}
\usepackage[left=23mm,right=13mm,top=35mm,columnsep=15pt]{geometry} 
\usepackage{adjustbox}
\usepackage{placeins}
\usepackage[T1]{fontenc}
\usepackage{lipsum}
\usepackage{csquotes}
\usepackage{siunitx}
\usepackage{textcomp}
\usepackage{ragged2e}

\usepackage{multirow}

\usepackage{titlesec}

\usepackage{tabularx}

\addto\captionsenglish{}
\addto\captionsenglish{}

%% file: sections/section01.tex
\section{INTRODUCTION} \label{sec:sec1}
Despite extensive studies of the charmonium system since its discovery in 1974~\cite{Aubert:1974js, Augustin:1974xw, Bacci:1974za, Abrams:1974yy}, knowledge of the singlet state $h_c$(1$^1P_1$) is sparse. Only nine decay modes have been observed, with the $h_c \rightarrow \gamma \eta_c$ predominant, with a branching fraction of (50$~\pm~$9)\%~\cite{PDG}. The $\psi(2S)$ to $h_c$(1$^1P_1$) hadronic transition is also known with a relatively large uncertainty, its branching ratio being (8.6$~\pm~$1.3)$\times 10^{-4}$~\cite{PDG}. \\ Although several measurements of its mass have been performed by the BESIII~\cite{PhysRevLett.104.132002, PhysRevD.86.092009}, the CLEO~\cite{PhysRevLett.101.182003}, and the E835~\cite{PhysRevD.72.032001} collaborations, a better precision is desirable. This would allow further tests of the hypothesis of zero spin-spin hyperfine mass splitting relative to the center-of-gravity mass of the three $\chi_{cJ}$(1$^3P_J$) states, defined as~\cite{HQS}
\begin{equation*}
M({\rm c.o.g.}) = \frac{M(\chi_{c0}) + 3M(\chi_{c1}) + 5M(\chi_{c2})}{9},
\end{equation*}
\noindent
where the $M(\chi_{cJ})$ are the masses of the $\chi_{cJ}$(1$^3 P_J$) states. Finally, only one measurement of the $h_c$(1$^1P_1$) width exists, provided by the BESIII experiment based on 106 million $\psi(2S)$ events~\cite{PhysRevD.86.092009}. The dataset used in the current analysis includes that sample. \\

A more precise knowledge of the $h_c$(1$^1P_1$) resonance parameters is also important given the recent discoveries in the $XYZ$ (charmonium-like) sector with the $h_c$ resonance as an intermediate state. Indeed, BESIII observed the $Z^{\pm}_c$(4020) decaying to $\pi^{\pm} h_c$~\cite{PhysRevLett.111.242001} as well as two resonant structures in the cross section for $e^+e^- \rightarrow \pi^+\pi^- h_c$~\cite{PhysRevLett.118.092002}. The signal decays of these exotic states all employ the tagged channel $h_c \rightarrow \gamma \eta_c, \eta_c \rightarrow {\rm hadrons}$. Thus, decreasing the uncertainty on the $h_c$ branching fractions and resonance parameters is of the utmost importance.\\

This article reports an improved determination of the $h_c$(1$^1 P_1$) resonance parameters taking advantage of $(448.1~\pm~2.9)\times~10^{6}$ $\psi(2S)$ collected by the BESIII detector in 2009 and 2012~\cite{Ablikim_2018}. The mass and width are extracted by studying the $\psi(2S) \rightarrow \pi^0 h_c \rightarrow (\gamma \gamma) (\gamma \eta_c)$ process following the same approach as Ref.~\cite{PhysRevLett.104.132002}. In particular, this study uses the $\pi^0$ recoil mass distribution to reconstruct the $h_c$(1$^1 P_1$) mass, both inclusively ($\psi(2S) \rightarrow \pi^0 h_c$) and by tagging the decay via the electric dipole (E1) transition photon from $h_c \rightarrow \gamma \eta_c$. The $\pi^0$ recoil mass ($RM(\pi^0)$) is defined as follows:
\begin{equation*}
RM(\pi^0) = \sqrt{(E_{\psi(2S)} - E_{\pi^0})^2 - (\bar{p}_{\psi(2S)} - \bar{p}_{\pi^0})^2} ,
\end{equation*}
\noindent
where $E_{\pi^0}$ ($E_{\psi(2S)}$) and $\bar{p}_{\pi^0}$ ($\bar{p}_{\psi(2S)}$) are the $\pi^0$ ($\psi(2S)$) energy and the momentum in the reference frame of the laboratory. \\

For the rest of this article, the two data samples will be referred as inclusive and tagged (Inc or Tag as a subscript), respectively. The choice of using two channels is motivated by the necessity of measuring the branching fractions $\mathcal{B}_{\rm Inc}(\psi(2S) \rightarrow \pi^0h_c)$ and $\mathcal{B}_{\rm Tag}(h_c \rightarrow \gamma \eta_c)$, the uncertainties of which are still large~\cite{PDG}. \\

%% file: sections/section02.tex
\section{BESIII DETECTOR AND DATASETS} \label{sec:DecandData}
The BESIII detector~\cite{ABLIKIM2010345} records symmetric $e^+e^-$ collisions provided by the BEPCII storage ring~\cite{Yu:IPAC2016-TUYA01}, which operates in the center-of-mass energy range from 2.0 to 4.9$\,$GeV. BESIII has collected large data samples in this energy region~\cite{Ablikim:2019hff}. The cylindrical core of the BESIII detector covers 93\% of the full solid angle and consists of a helium-based multilayer drift chamber (MDC), a plastic scintillator time-of-flight system (TOF), and a CsI(Tl) electromagnetic calorimeter (EMC), which are all enclosed in a superconducting solenoidal magnet providing a 1.0$\,$T magnetic field. The solenoid is supported by an octagonal flux-return yoke with resistive plate counter muon identification modules interleaved with steel. The charged-particle momentum resolution at 1$\,{\rm GeV}/c$ is $0.5\%$, and the $dE/dx$ resolution is $6\%$ for electrons from Bhabha scattering. The EMC measures photon energies with a resolution of $2.5\%$ ($5\%$) at 1$\,$GeV in the barrel (end cap) region. The time resolution in the TOF barrel region is 68$\,$ps, while in the end cap is 110$\,$ps. \\

Simulated data samples produced with a {\sc geant4}-based~\cite{Agostinelli:2002hh} Monte Carlo (MC) package, which includes the geometric description of the BESIII detector and the detector response, are used to determine detection efficiencies and to estimate backgrounds. The simulation models the beam energy spread and initial state radiation (ISR) in the $e^+e^-$ annihilations with the generator {\sc kkmc}~\cite{PhysRevD.63.113009, Jadach:1999vf}. The inclusive MC sample consists of the production of the $\psi$(2S) resonance, the ISR production of the $J/\psi$, and the continuum processes ($e^+e^- \rightarrow e^+e^-$, $e^+e^- \rightarrow$ hadrons, and $e^+e^- \rightarrow \gamma\gamma$) incorporated in {\sc kkmc}. The known decay modes are modelled with {\sc evtgen}~\cite{Lange:2001uf, 2007-0205} using branching fractions taken from the Particle Data Group (PDG)~\cite{PDG}, and the remaining unknown charmonium decays are modelled with {\sc lundcharm}~\cite{PhysRevD.62.034003, Yang_2014}. Final state radiation (FSR) from charged final state particles is incorporated using the {\sc photos}~\cite{photos} package. In the signal MC samples, the $\psi(2S) \rightarrow \pi^0 h_c$ decay and the E1 transition $h_c \rightarrow \gamma \eta_c$ are both generated by {\sc evtgen}. 

%% file: sections/section03.tex
\section{EVENT SELECTION AND ANALYSIS PROCEDURE} \label{sec:sec3}
Although charged tracks are not directly used in the reconstruction, candidate events must have at least two good tracks to suppress background.  Good tracks reconstructed in the MDC must pass the following fiducial and production vertex cuts.  Only tracks with momenta less than 2.0$\,$GeV/$c$ are considered and they are required to satisfy \mbox{$|\rm{cos\theta}|$ < 0.93}, where $\theta$ is the angle between the momentum and beam axis. 
The distance of closest approach to the interaction point must be less than 10$\,$cm along the beam axis, and less than 1$\,$cm in the transverse plane. 
 Photon candidates are identified using showers in the EMC. The deposited energy of each shower must be more than 25$\,$MeV in the barrel region ($|\rm{cos\theta}|$ < 0.80) and more than 50$\,$MeV in the end cap region (0.86 < $|\rm{cos\theta}|$ < 0.92). To suppress electronic noise and showers unrelated to the event, the difference between the EMC time and the event start time is required to be within (0, 700)$\,$ns.  To exclude showers that originate from charged tracks, the angle between the position of each shower in the EMC and the closest extrapolated charged track must be greater than 10$^\circ$. Finally, to suppress the background contribution from the continuum processes, such as Bhabha scattering, the total energy deposit in the EMC ($E^{\rm Tot} _{\rm EMC}$) is required to satisfy \mbox{0.6 < $E^{\rm Tot} _{\rm EMC}$ < 3.2$\,$GeV}. Candidate events are required to have at least two (three) photon candidates passing the fiducial and energy cuts for the inclusive (tagged) channel. The $\pi^0$ candidate is reconstructed via its decay to $\gamma \gamma$, with both photon candidates laying in the barrel ($|\rm{cos\theta}|$ < 0.80) and having an energy $\geq$ 40$\,$MeV. The di-photon pair is accepted as a $\pi^0$ candidate if its invariant mass satisfies \mbox{120 < $M_{\gamma \gamma}$ < 145$\,$MeV/$c^2$}. Multiple $\pi^0$ candidates are allowed in one event.  A one-constraint (1C) kinematic fit fixing the $\pi^0$ mass to its known value~\cite{PDG} is used to improve the energy resolution.  Candidates with a fit $\chi^2 > 200$ are rejected. 
The dominant background for the inclusive sample comes from combining photons from two different $\pi^0$s; the shape of this background is discussed below. The photon coming from the E1 decay ($\gamma_{\rm Tag}$) is expected to peak around 500$\,$MeV. Taking into consideration that $\Gamma_{\eta_c} \approx	30\,$MeV~\cite{PDG}, the E1 photon is required to satisfy \mbox{465 < $E\gamma_{\rm{Tag}}$ < 535$\,$MeV} and must not form a $\pi^0$ with any other photons in the event. In the tagged channel, if more than one $\pi^0$ is found in the $\pi^0$ recoil mass signal region (3.500--3.550$\,$GeV/$c^2$) the $\pi^0$ with the minimum 1C fit $\chi^2$ is kept. \\

The three main background sources are $\psi(2S)\rightarrow J/\psi \, \pi^+ \pi^-$, $\psi(2S)\rightarrow J/\psi \, \pi^0 \pi^0$, and $\psi(2S)\rightarrow\gamma \chi_{c0}$. The first two are suppressed by requiring all combinations of the $\pi^+ \pi^-$ ($\pi^0 \pi^0$) recoil mass to be outside the range $M_{J/\psi} \pm 4$ MeV/$c^2$ ($M_{J/\psi} \, ^{+38} _{-8}$ MeV/$c^2$), where $M_{J/\psi}$ is the nominal ${J/\psi}$ mass~\cite{PDG}. This veto window is optimized based on the figure of merit $\frac{\rm{S}}{\sqrt{\rm{S+B}}}$, where S and B are the numbers of signal and background events estimated from the MC. This selection allows to remove $\sim76\%$ ($\sim96\%$) of the $\psi(2S)\rightarrow J/\psi \, \pi^+ \pi^-$ ($\psi(2S)\rightarrow J/\psi \, \pi^0 \pi^0$) background events, while cutting less than $5\%$ ($1\%$) of the signal events. No suppression of the $\gamma \chi_{c0}$ decay is needed since the photon energy is not in the tagged photon energy range, and thus this decay contributes to the background with a typical combinatorial shape. \\

\begin{figure}[!h]
		\centering
			\subfigure{
				\includegraphics[width=0.5\textwidth]{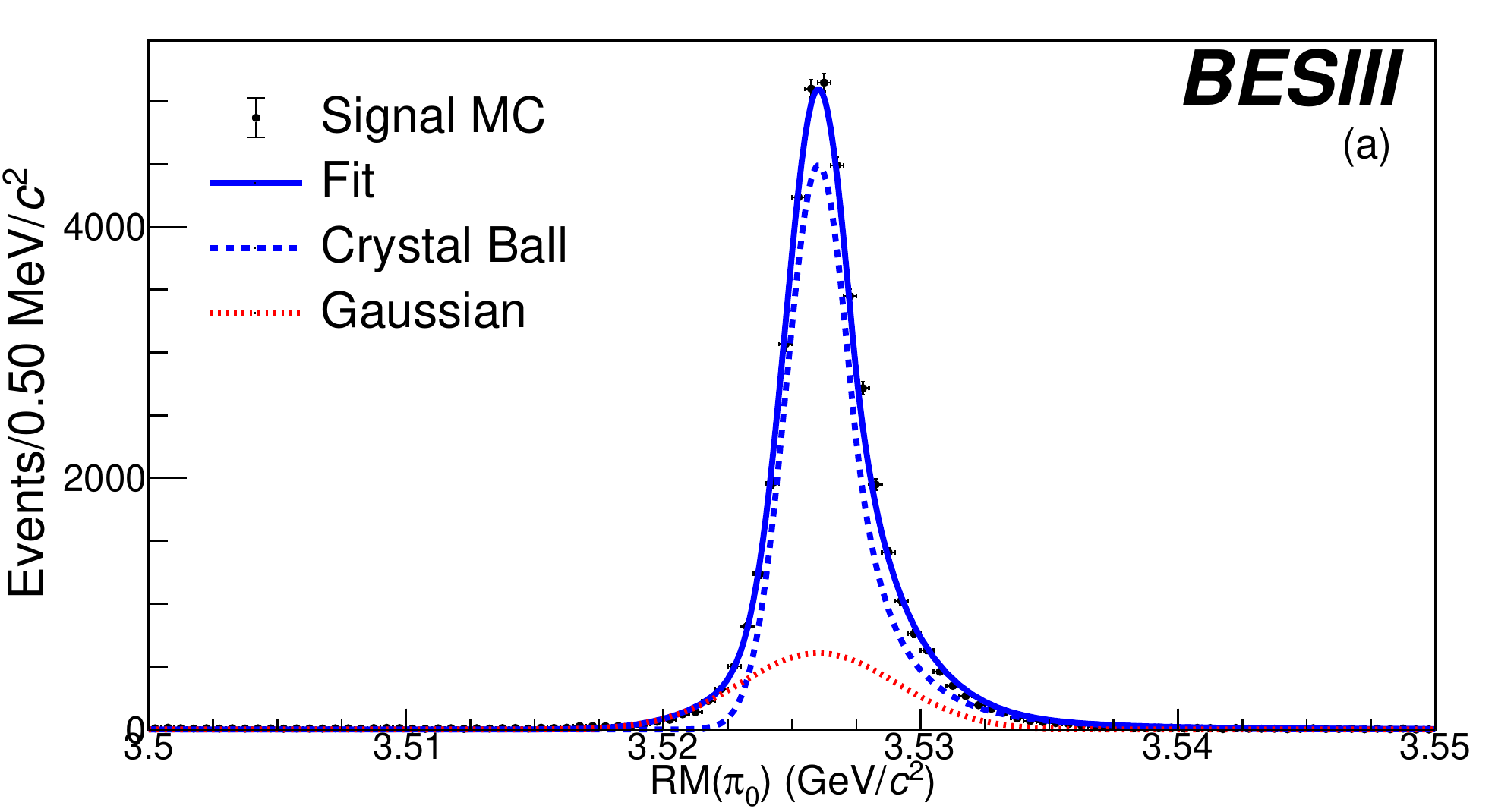}
		}
		\subfigure{
	   		\includegraphics[width=0.5\textwidth]{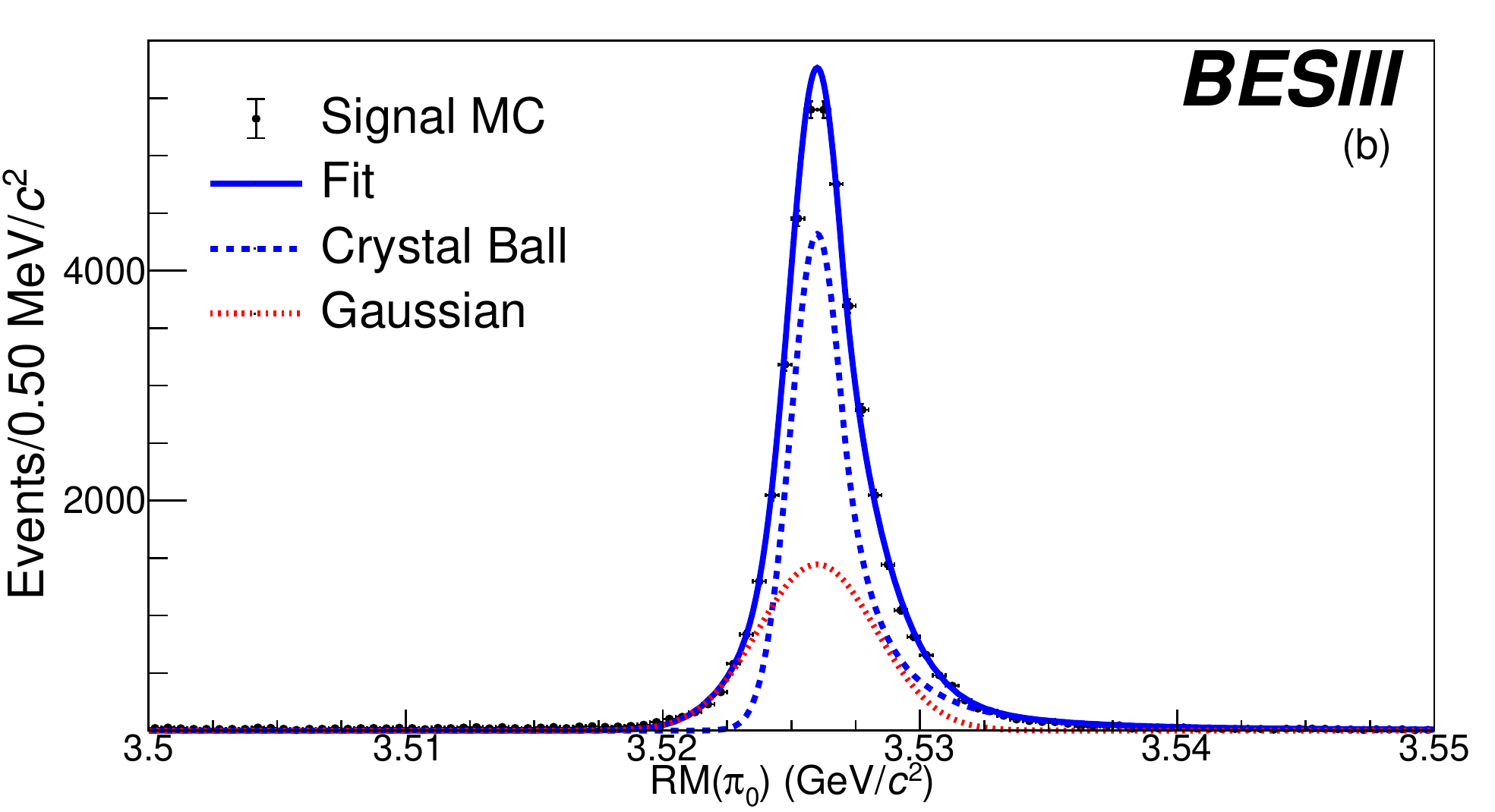}
		  }
	\caption[]{Signal MC $\pi_0$ recoil mass (referred as RM($\pi_0$)) distributions for the $h_c$(1$^1P_1$) with its width set to 0. The solid blue line refers to the global fit result, while the black dots with error bars are the MC simulation. The blue dashed line represents the Crystal Ball component, while the red dotted one is the Gaussian component. Fig.~(a) shows the resolution function for the inclusive channel, while the one for the tagged chain is presented in Fig.~(b).}
	\label{fig:reso_fnc}
\end{figure}

The signal efficiencies, $\epsilon_{\rm{Tag}}$ = 12.37\% and $\epsilon_{\rm{Inc}}$ = 14.25\%, are estimated for the tagged and inclusive channels, respectively, based on two signal MC samples of 300000 events each.  
The signal shape is modelled with a resolution function (shown in Fig.~\ref{fig:reso_fnc}) based on MC.  
It is a sum of Gaussian and Crystal Ball functions; both contribute to symmetric smearing while the latter also includes the low-energy tail due to $\pi^0$ reconstruction. The parameters of this resolution function are obtained from a signal MC dataset where the $h_c$(1$^1P_1$) width is set to 0. The convolution of the resolution function and a Breit-Wigner distribution is used to describe the signal in the final dataset to extract $N\rm{_{\rm Inc}}$ and $N\rm{_{\rm Tag}}$ (i.e., the number of the inclusive and tagged events, respectively). Defining $N(\psi$(2S)) as the total number of $\psi$(2S) events, and $\mathcal{B}(\pi^0 \rightarrow \gamma \gamma)$ as the branching fraction of the $\pi^0 \rightarrow \gamma \gamma$ decay (taken from PDG~\cite{PDG}), the various branching fractions, $\mathcal{B}$, are obtained as

\begin{itemize}
 \item $\mathcal{B}_{\rm Inc}(\psi(2S)\rightarrow\pi^0h_c)\times\mathcal{B}_{\rm Tag}(h_c\rightarrow\gamma\eta_c)=\frac{N_{{\rm Tag}}}{\epsilon_{{\rm Tag}} \times N(\psi(2S)) \times \mathcal{B}(\pi^0 \rightarrow \gamma \gamma) }$; 

\item $\mathcal{B}_{\rm Inc}(\psi(2S)\rightarrow\pi^0h_c)=\frac{{N_{\rm Inc}}}{\epsilon_{{\rm Inc}}\times N(\psi(2S)) \times \mathcal{B}(\pi^0 \rightarrow \gamma \gamma) }$;

\item $\mathcal{B}_{\rm Tag}(h_c\rightarrow\gamma\eta_c)=\frac{\mathcal{B}_{\rm Inc}(\psi(2S)\rightarrow\pi^0h_c)\times\mathcal{B}_{\rm Tag}(h_c\rightarrow\gamma \eta_c)}{\mathcal{B}_{\rm Inc}(\psi(2S)\rightarrow\pi^0 h_c)}=\frac{N_{\rm Tag}\times\epsilon_{\rm Inc}}{N_{\rm Inc}\times\epsilon_{\rm Tag}}$.
\end{itemize}

To assess the shape of the irreducible background, a MC dataset (described in Sec.\,\ref{sec:DecandData}) of 400 million inclusive $\psi(2S)$ decays with the signal contributions removed is studied. 
This shows that a 5$^{\rm{th}}$-order Chebychev polynomial satisfactorily describes the irreducible background in both channels. \\

Fits to the distributions of the $\pi^0$ recoil mass are performed minimizing the negative log-likelihood on the inclusive and tagged channels separately. The total probability function is constructed from a 5$^{\rm{th}}$-order Chebychev polynomial, to describe the background, added to a convolution of the resolution function and a Breit-Wigner distribution for the signal. In the tagged channel, the parameters of the Chebychev polynomial are allowed to float, as well as the mass and the width of the $h_c$(1$^1P_1$) resonance.  But for the inclusive channel, the signal shape parameters are fixed to the values found in the tagged channel, while the parameters of the Chebychev polynomial are left floating. Input-output tests on MC samples are used to validate both the model and the fitting procedure; no bias is found. 

\begin{figure}[!h]
		\centering
			\subfigure{
   				\includegraphics[width=0.5\textwidth]{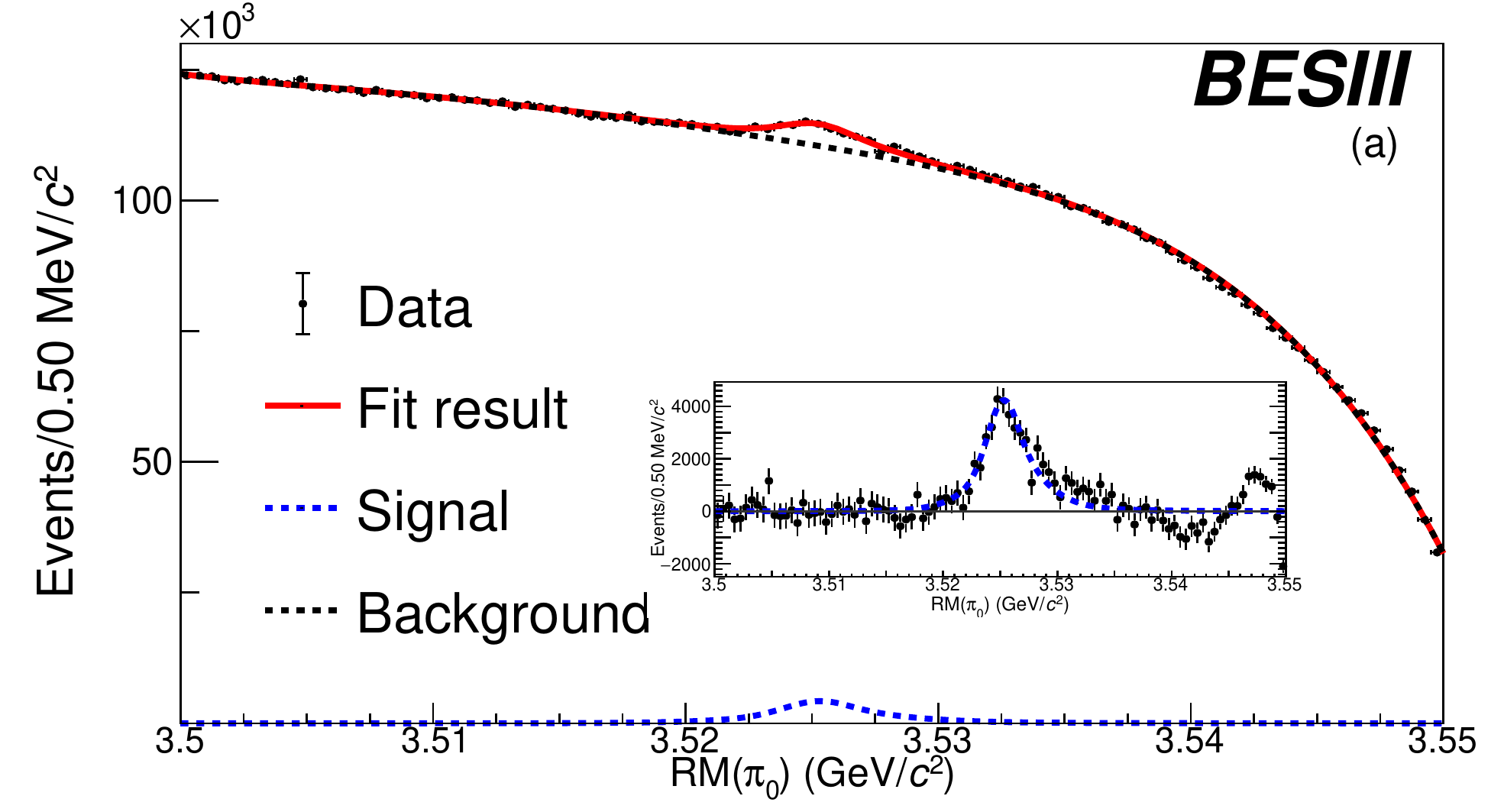}
			}
			\subfigure{
		   		\includegraphics[width=0.5\textwidth]{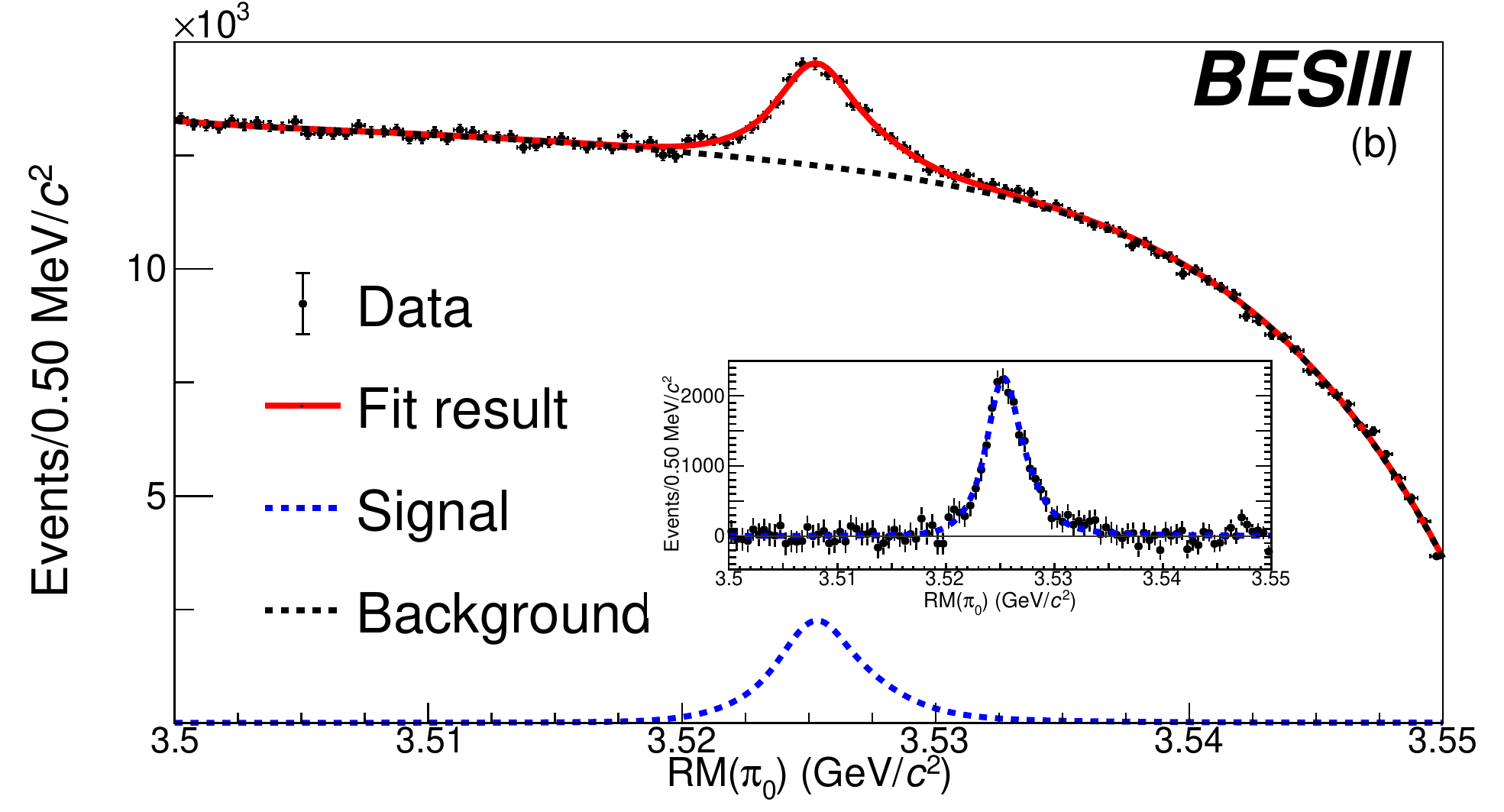}
	 		  }
		\caption[]{Fits to the $\pi^0$ recoil mass spectra for the (a) inclusive and (b) tagged samples. Red solid lines denote the fit results, while the black dots with error bars are the data. The blue dashed lines represent the signal component and the black dashed lines are the background. Insets show the background-subtracted data with the signal shape overlaid.} 
		\label{fig:Fittot}
	\end{figure}

The fit results are summarized in Table~\ref{tab:the_Fit}. A comparison between this analysis and the PDG\cite{PDG} shows the compatibility between the central values with improved precision. Fig.~\ref{fig:Fittot} presents the fits for the $\pi^0$ recoil mass spectra, RM($\pi_0$), for both the inclusive and tagged channels. 
Separate fits to the 2009 and 2012 datasets yield parameters compatible with each other within 1.0$\sigma$. 
The systematic uncertainties reported in Table~\ref{tab:the_Fit} are described in detail in the next section. 

\begin{table}[htbp]
\centering
\caption[]{Results of the fits to the $\pi^0$ recoiling mass spectra with statistical (first) and systematic (second) uncertainties. The last column provides current PDG values.}
\begin{tabular}{ccc}
\toprule
Variable & Value & PDG Value~\cite{PDG}\\
\colrule 
$M(h_{c})$ (MeV/$c^{\rm{2}}$) & $3525.32 \pm 0.06 \pm 0.15$ & 3525.38 $\pm$ 0.11 \\
%${\mathit \Gamma}(h_{c})$ (MeV) & 0.78 $^{+0.27}_{-0.24}$ $\pm$ 0.12 & 0.7 $\pm$ 0.4 \\
\multirow{2}{*}[1ex]{\centering ${\mathit \Gamma}(h_{c})$ (MeV) }& \multirow{2}{*}[1ex]{\centering $0.78^{+0.27}_{-0.24} \pm 0.12$} & 0.70 $\pm$ 0.28 $\pm$ 0.22 \\[-3mm] 
& & {\tiny (BESIII \cite{PhysRevD.86.092009})} \\
$N_{\rm{Tag}}(h_{c})$ & 23118 $^{+1500}_{-1398}$ & --- \\
\multirow{4}{*}[2ex]{\centering $\mathcal{B}_{\rm Inc} \times \mathcal{B}_{\rm Tag}$ (10$^{-4}$) }& \multirow{4}{*}[2ex]{\centering $4.22 ^{+0.27}_{-0.26} \pm 0.19$} & $4.58\pm0.64$ \\[-3mm] 
& & {\tiny (BESIII \cite{PhysRevLett.104.132002})} \\[-1mm] 
& & $4.16 \pm 0.48$ \\[-3mm] 
& & {\tiny (CLEO \cite{PhysRevLett.101.182003})}\\
$N_{\rm{Inc}}(h_{c})$ & 46187 $\pm$ 2123 & --- \\
%$\mathcal{B}_{\rm Inc}$ (10$^{-4}$) & $7.23 \pm 0.33 \pm 0.38$ & 8.60 $\pm$ 1.30 \\
\multirow{4}{*}[2ex]{\centering $\mathcal{B}_{\rm Inc}$ (10$^{-4}$)  }& \multirow{4}{*}[2ex]{\centering $7.32 \pm 0.34 \pm 0.41$} & $8.40 \pm 1.30 \pm 1.00$ \\[-3mm] 
& & {\tiny (BESIII \cite{PhysRevLett.104.132002})} \\[-1mm] 
& & $9.00 \pm 1.5 \pm 1.3$ \\[-3mm] 
& & {\tiny (CLEO \cite{PhysRevD.84.032008})}\\
%$N_{\rm{Inc}}(h_{c})$ & 46187 $\pm$ 2123 & --- \\
%$\mathcal{B}_{\rm Tag}$ (\%) & $57.66 ^{+3.62}_{-3.50}\pm 0.58$ & 50 $\pm$ 9 \\
\multirow{4}{*}[2ex]{\centering $\mathcal{B}_{\rm Tag}$ (\%) } & \multirow{4}{*}[2ex]{\centering $57.66 ^{+3.62}_{-3.50} \pm 0.58$} & $53 \pm 7 \pm 8$ \\[-3mm] 
& & {\tiny (BESIII \cite{PhysRevLett.104.132002})} \\[-1mm] 
& & $48 \pm 6 \pm 7$ \\[-3mm] 
& & {\tiny (CLEO \cite{PhysRevLett.101.182003})}\\
\botrule
\end{tabular}\\[10pt]
\label{tab:the_Fit}
\end{table}

%% file: sections/section04.tex
\section{SYSTEMATIC UNCERTAINTIES} 
\label{sec:sec4}

Sources of systematic uncertainties on the measurement of the $h_c$(1$^1P_1$) resonance parameters and branching fractions include the background line-shape, the mass ranges for  the veto of $J/\psi$ from $\psi(2S)$, the photon energy calibration and reconstruction efficiency, the $\pi^0$ reconstruction efficiency, and the luminosity. The contributions from each source are shown in Table~\ref{tab:sys}. For each measurement, the total systematic uncertainty corresponds to a quadrature sum of all individual sources, which are discussed in detail next. \\

\paragraph*{Background line-shape.} To determine the systematic uncertainties associated to the fit function, a 4$^{\rm th}$-order Chebychev polynomial to describe the background behaviour in both the inclusive and tagged channels is tested. The discrepancy with respect to the nominal fit result is taken as a systematic uncertainty. 
Assuming that at the $\pi^0$ reconstruction level (due to the fact that the $\eta_c$ is left to decay inclusively) some background events might be miscounted as signal, a first-order polynomial is added to the resolution function, maintaining a 4$^{\rm th}$ order Chebychev polynomial to describe the main background contribution. The discrepancy between this method and the nominal one is taken as a systematic uncertainty and summed in quadrature with the result from the first variation. \\

\paragraph*{Mass ranges for the veto of $J/\psi$ from $\psi(2S)$.} Mass veto ranges for the $\pi^0 \pi^0$ and $\pi^+ \pi^-$ recoiling masses are changed according to the observed variations of the figure of merit. Two different scenarios are tested,

\begin{itemize}
    \item for the $\pi^+$ $\pi^- J/\psi$ background, $M_{J/\psi} \pm$ 3 MeV/$c^{\rm{2}}$ and $M_{J/\psi} \pm$ 5 MeV/$c^{\rm{2}}$;
    \item for the $\pi^0$ $\pi^0 J/\psi$ background, $M_{J/\psi}$$^{+37}_{-7}$ MeV/$c^{\rm{2}}$  and $M_{J/\psi}$$^{+39}_{-9}$ MeV/$c^{\rm{2}}$ .
\end{itemize}

\noindent
In both cases, smaller veto ranges give the largest changes with respect to the nominal vetoes; these changes are taken as systematic uncertainties. \\

\paragraph*{Photon reconstruction efficiency.} The systematic uncertainties arising from potential inconsistencies of the photon-energy measurements between data and MC simulation are obtained from Ref.~\cite{PhysRevD.86.092009}. \\ 

\paragraph*{Photon energy calibration.} The uncertainty due to the photon energy distribution is obtained from Ref.~\cite{PhysRevD.86.092009}. \\ 

\paragraph*{Signal shape.} This uncertainty includes contributions from the signal line-shape and the 1-C kinematic fit, and is estimated in Ref.~\cite{PhysRevD.86.092009}. \\

\paragraph*{$\pi^0$ reconstruction efficiency.} The uncertainty on the branching fractions due to the $\pi^0$ reconstruction efficiency is estimated in Ref.~\cite{PhysRevD.81.052005}. \\ 

\paragraph*{Number of $\psi(2S)$ events.} This uncertainty is estimated in Ref.~\cite{Ablikim_2018}, and included in the branching fraction uncertainties. \\ 

Other sources of the possible systematic uncertainties are studied, but found to be negligible. These include the bin size, the $\pi^0$ recoil mass signal region, the trigger efficiency, the numbers of $\pi^0$ and charged tracks, the masses and widths of the $\psi(2S)$ and $\eta_c$, and the sample sizes of the MC simulations. \\

\begin{table}[t]
	\centering
		\caption[]{Summary of the relative systematic uncertainties on the $h_c$(1$^1P_1$) resonance parameters and branching fractions. The --- symbol is used when the systematic sources are found to be negligible or not applicable. For each measurement, the total systematic uncertainty corresponds to a quadrature sum of all individual contributions.}
		\begin{tabular}{cccccc}
                    \toprule
                    \multirow{2}{*}[0ex]{\centering Source} & \multirow{2}{*}[1ex]{\centering $M(h_c)$} & \multirow{2}{*}[1ex]{\centering  ${\mathit \Gamma}(h_c)$} & \multirow{2}{*}[1ex]{\centering $\mathcal{B}_{\rm Inc} \times \mathcal{B}_{\rm Tag}$} & \multirow{2}{*}[1ex]{\centering $\mathcal{B}_{\rm Inc}$} & \multirow{2}{*}[1ex]{$\mathcal{B}_{\rm Tag}$} \\ 
		   & {\small ($10^{-3}$)} & ($10^{-2}$)  & ($10^{-2}$)  & ($10^{-2}$)  & ($10^{-2}$) \\
                    %Source & $M(h_c)$ & ${\mathit \Gamma}(h_c)$ & $\mathcal{B}_{\rm Inc} \times \mathcal{B}_{\rm Tag}$ & $\mathcal{B}_{\rm Inc}$ & $\mathcal{B}_{\rm Tag}$ \\
		   \colrule 
		   Background & \multirow{2}{*}{0.02} & \multirow{2}{*}{---} & \multirow{2}{*}{0.44} & \multirow{2}{*}{2.86} & \multirow{2}{*}{---} \\[-2mm] shape & & & & & \\ 
		    $J/\psi$ veto & 0.01 & 9.33 & 3.17 & 4.15 & --- \\
		    $\gamma$ reconstruction & \multirow{2}{*}{---} & \multirow{2}{*}{---} & \multirow{2}{*}{3.00} & \multirow{2}{*}{2.00} & \multirow{2}{*}{1.00} \\[-2mm] efficiency & & & & & \\
		   $\gamma$ energy & \multirow{2}{*}{0.04} & \multirow{2}{*}{8.97} & \multirow{2}{*}{---} & \multirow{2}{*}{---} & \multirow{2}{*}{---} \\[-2mm] calibration & & & & & \\
		   Signal shape & --- & 7.69 & --- & --- & --- \\
   		   $\pi^0$ reconstruction & \multirow{2}{*}{---} & \multirow{2}{*}{---} & \multirow{2}{*}{1.00} & \multirow{2}{*}{1.00} & \multirow{2}{*}{---} \\[-2mm] efficiency & & & & & \\ 
		   Number of & \multirow{2}{*}{---} & \multirow{2}{*}{---} & \multirow{2}{*}{0.65} & \multirow{2}{*}{0.65} & \multirow{2}{*}{---} \\[-2mm] $\psi(2S)$ events & & & & & \\
		   Total systematic & \multirow{2}{*}{0.04} & \multirow{2}{*}{15.06} & \multirow{2}{*}{4.55} & \multirow{2}{*}{5.55} & \multirow{2}{*}{1.00} \\[-2mm] uncertainty & & & & & \\
		   \botrule
               \end{tabular}
		\label{tab:sys}
\end{table}

 \multirow{2}{*}{}

%% file: sections/section05.tex
\section{RESULTS AND SUMMARY} \label{sec:sec5}

In this article, two decay chains involving the $h_c(1^1 P_1)$ charmonium state are studied,\\

\begin{itemize}
\item inclusive: $\psi(2S) \rightarrow \pi^0 h_c$ with $h_c \rightarrow {\rm anything}$,
\item tagged: $\psi(2S) \rightarrow \pi^0 h_c$ with $h_c \rightarrow \gamma \eta_c$.
\end{itemize}

The measurements of the $h_c(1^1P_1)$ resonance parameters and branching fractions are performed with the world's largest $\psi(2S)$ data sample, collected by the BESIII experiment~\cite{Ablikim_2018}. This work provides the second estimate of the $h_c(1^1P_1)$ width, given with a similar uncertainty as a previous BESIII measurement which used 16 exclusive decays to reconstruct the $\eta_c$ state~\cite{PhysRevD.86.092009}. The novelty provided by this analysis resides in the increased branching fractions precision, which is improved between two and three times with respect to the PDG measurements~\cite{PDG}. Despite the decreased uncertainties, this work's branching ratios remain compatible with both CLEO~\cite{PhysRevLett.101.182003, PhysRevD.84.032008} and BESIII~\cite{PhysRevLett.104.132002} estimates. All the discussed features, summarised in Table~\ref{tab:the_Fit} altogether with the PDG ones~\cite{PDG}, are compatible with the world values~\cite{PDG} within one standard deviation. \\

Furthermore with the $h_c(1^1P_1)$ mass estimate of this work, using current PDG~\cite{PDG} values for the center-of-gravity mass of the three $\chi_{cJ}$(1 $^3P_J$) states, $M(\rm c.o.g.)$, an updated value for the 1$P$ hyperfine mass splitting (\mbox{$\Delta_{\rm hyp} = M(h_c) - M({\rm c.o.g.})$}) may be obtained. This value is predicted by QCD to be identical to 0 at leading-order~\cite{HQS}. No mass splitting is observed with this measurement,
\begin{equation*}
\Delta_{\rm hyp} = 0.03 \pm 0.06 ~ (\rm{stat.}) \pm ~ 0.15 ~ (\rm{syst.}) ~ \rm{MeV}/c^{\rm{2}}.
\end{equation*}
These results on $\Delta_{\rm hyp}$ show that, with the foreseen increase in $\psi(2S)$ statistics by the BESIII experiment~\cite{Ablikim:2019hff}, significant efforts to reduce systematic uncertainties will be necessary. 

%% file: sections/acknowledgements.tex
\section*{Acknowledgements} \label{sec:acknowledgements}

The BESIII collaboration thanks the staff of BEPCII and the IHEP computing center for their strong support. This work is supported in part by National Key R$\&$D Program of China under Contracts Nos. 2020YFA0406300, 2020YFA0406400; National Natural Science Foundation of China (NSFC) under Contracts Nos. 11625523, 11635010, 11735014, 11822506, 11835012, 11935015, 11935016, 11935018, 11961141012, 12022510, 12025502, 12035009, 12035013, 12061131003; the Chinese Academy of Sciences (CAS) Large-Scale Scientific Facility Program; Joint Large-Scale Scientific Facility Funds of the NSFC and CAS under Contracts Nos. U1732263, U1832207; CAS Key Research Program of Frontier Sciences under Contract No. QYZDJ-SSW-SLH040; 100 Talents Program of CAS; INPAC and Shanghai Key Laboratory for Particle Physics and Cosmology; ERC under Contract No. 758462; European Union Horizon 2020 research and innovation programme under Contract No. Marie Sklodowska-Curie grant agreement No 894790; German Research Foundation DFG under Contracts Nos. 443159800, Collaborative Research Center CRC 1044, FOR 2359, FOR 2359, GRK 214; Istituto Nazionale di Fisica Nucleare, Italy; Ministry of Development of Turkey under Contract No. DPT2006K-120470; National Science and Technology fund; Olle Engkvist Foundation under Contract No. 200-0605; STFC (United Kingdom); The Knut and Alice Wallenberg Foundation (Sweden) under Contract No. 2016.0157; The Royal Society, UK under Contracts Nos. DH140054, DH160214; The Swedish Research Council; U. S. Department of Energy under Contracts Nos. DE-FG02-05ER41374, DE-SC-0012069.